# UPGRADE OF THE ILC CRYOMODULE

A. Basti[4], F. Bedeschi[4], A. Bryzgalin[5], J. Budagov[1], P. Fabbricatore[4], E. Harms[3], S. Illarionov[5], S. Nagaitsev[3], E. Pekar[5], V. Rybakov[2], B. Sabirov[1*], Ju. Samarokov[2], G. Shirkov[1], W. Soyars[3], Ju. Taran[1], G. Trubnikov[1]

1-Joint Institute for Nuclear Research, Dubna, Russia; 2- Russian Federal Nuclear Center–All-Russian Research Institute of Experimental Physics, Sarov, Russia; 3- Fermi National Accelerator Laboratory, Batavia, USA; 4- Istituto Nazionale di Fisica Nucleare, Pisa/Genova, Italy; 5- Paton Electric Welding Institute, Kiev, Ukraine
*- sabirov@jinr.ru

## INTRODUCTION

In May 2005, the International Committee on Future Accelerators (ICFA) at the International Union of Pure and Applied Physics (IUPAP) claimed achievement of consensus and consequent necessity to globalize and unite efforts of world's research centers for building a new-generation accelerator complex, that is, an electron–positron linear collider. The current universally adopted title of this unique 21st century project is the International Linear Collider (ILC).

The ILC is planned to allow electron–positron collisions at energies of 500 to 1000 GeV and higher. It is the next step after the LHC, the world's largest particle accelerator operating at CERN (Geneva).

At the Joint Institute for Nuclear Research, the research on the ILC is included in the Topical Plan for JINR Research and International Cooperation.

This paper summarizes the ILC R@D results obtained by the JINR team in cooperation with the teams from FNAL (Batavia, US), INFN (Pisa/Genova, Italy), RFNC–VNIIEF (Sarov, Russia), and EWI (Kiev, Ukraine).

The accelerator is intended for accelerating electron–positron beams to the energy of 0.5 TeV (stage 1). The total length of the accelerating section is ~35 km (Fig. 1).

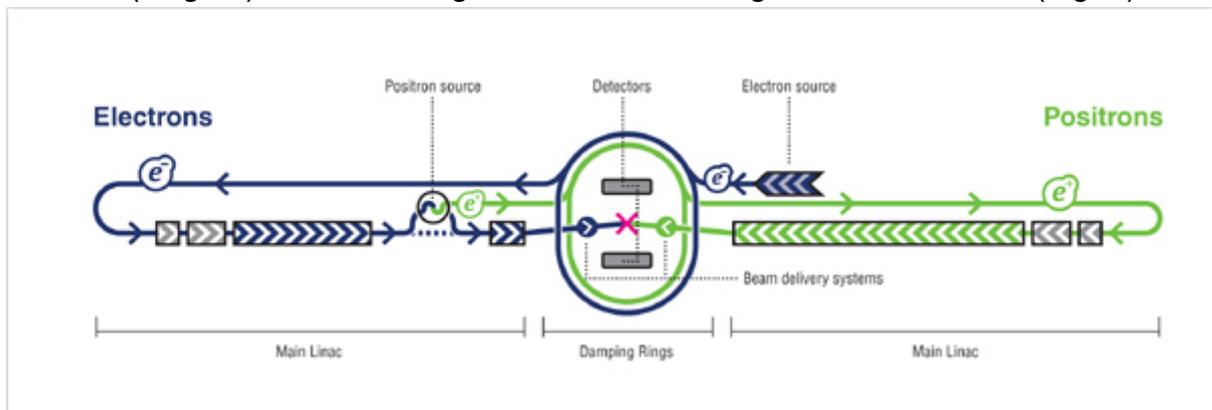

*Fig. 1. Schematic view of the ILC.*

Electrons and positrons are accelerated over this length with superconducting cavities of highly pure niobium (Nb). Their total number is about 20000. Since the cavities must be kept at a temperature of 1–2 K, they are placed in titanium (Ti) tanks filled with liquid helium. A titanium pipe is laid along the entire length of the accelerator to continuously supply cryostats with helium. The choice of titanium is dictated by specific welding properties of niobium and titanium.

## BIMETALLIC Ti+SS TRANSITION ELEMENTS

The administrative body of the ILC project has considered and JINR has investigated the stainless steel (SS) option of the helium supply pipe, which promised a substantial decrease in the project cost. The key problem of this option turned out to be transition from the stainless-steel pipe to the titanium tank (Fig. 2).

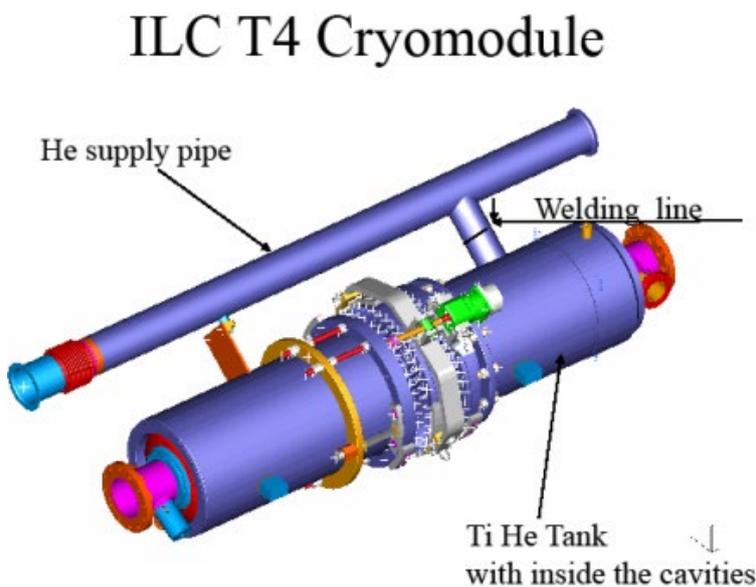

Fig. 2. Connection of the helium supply pipe to a titanium tank with a niobium cavity inside.

As is known, SS and Ti cannot be welded by a conventional method of electron beam welding. A group from KEK(Japan) tried to weld SS and Ti pieces using two nonconventional methods, friction welding and isostatic hot pressure welding. The pieces were in general welded, but the weld turned out to be as brittle as glass at ultralow operating temperatures of superconducting cryomodules. A more reliable welding method had to be searched for. It was found at RFNC–VNIIEF (Sarov) [1, 4, 9].

The unique welding method proposed in Sarov uses the explosion energy. Generally speaking, this method is known in the world, but it is used to weld flat pieces. In Sarov, they developed and practically used the method for welding cylindrical pieces. The following objectives were set:
* To develop a pilot technological explosion welding process for making a bimetallic tubular transition element.
* To investigate the microstructure of the welded joint.
* To test the welded joint for leakage at room and cryogenic temperatures.

The first bimetallic tubular transition element was made using Russian materials: 12X18H10T steel and BT1-0 titanium [2-3]. Metallographic investigations were carried with the following results:
1. No microdefects were found in the welded joints.

2. The welded joints revealed a sinusoidal character, which increased the joint strength. The wave length and amplitude were 0.3 mm and 0.05 mm respectively.

3. Explosion welding gave rise to a number of microdefects, but they were locally arranged and did not form a continuous layer.

4. Metal strengthening was observed in the welded joint area. It was most intensive in a narrow zone near the titanium–steel interface ~0.5 mm wide and decreased outside of it.

5. The shear strength test of the welded joint yielded an impressive value $τ_{sh}$ ≈ 250 MPa.

The sample was presented at the Conference on ILC in Milan in 2006, where it greatly impressed the participants and gained their unanimous approval.

The working version of the transition element was made of Chinese GRADE2 titanium and Austrian 316L stainless steel. Explosion welding was used to make a transition pipe using a band of Russian stainless steel (Fig. 3).

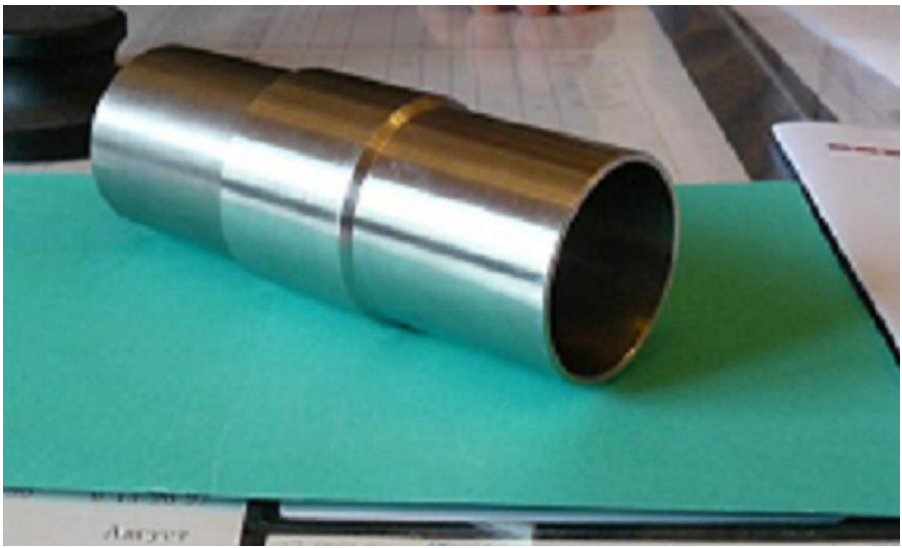

*Fig. 3. Pilot sample of a bimetallic pipe made of Chinese titanium and Austrian steel.*

This bimetallic sample was comprehensively tested in Dubna and Pisa [5-6].

1. In Dubna, the sample was subjected to six thermal test cycles of cooling with liquid nitrogen to 77 K and heating to 300 K. Leak-tightness was measured with a PFEIFER-VACUUM HLT160 leak detector at room temperature, and the result was Q≤$10^{-7}$ L·atm/s. Then a pressure of 6.5 atm was produced inside the sample using helium, and the sample was exposed to ultrasound at different temperatures and with different exposures. The leak detector reading remained the same, Q≤$10^{-7}$ L·atm/s. Next was the cryotemperature test which included six cycles of cooling the sample to ~6 K using the cryocooler and heating it to room temperature. Then leak-tightness was measured at room temperature, and the result was Q≤$10^{-7}$ L·atm/s.

2. In Pisa, thermal cycles with liquid nitrogen were repeated. Leak-tightness was measured with a Wis Technologies MODUL200 leak detector and found to be 1.2·$10^{-10}$ mbar·L/s at room temperature (300 K), 8.6·$10^{-9}$ mbar·L/s at the temperature of 200 K, and 3.4·$10^{-9}$ mbar·L/s at the temperature close to the liquid nitrogen temperature (83 K).

Upon agreement with the FNAL and INFN partners, it was decided to make ten bimetallic samples and repeat the comprehensive tests in Sarov, Dubna, Pisa, and Batavia to increase statistics of test results.

The samples were made in Sarov and subjected to multiple thermal cycling successively in Dubna and Pisa. The leak-tightness measurement results were as follows:

- a) at room temperature 300 K          $7.5 \cdot 10^{-10}$ Torr·L/s;
- b) at temperature 77 K                 $7.5 \cdot 10^{-9}$ Torr·L/s;
- c) at pressure 6.5 atm                 $<5 \cdot 10^{-10}$ Torr·L/s.

The next step in the investigation of the properties of the samples was their testing under real working conditions at FERMILAB: the pipes are supposed to be tested at the cryogenic temperature of 1.6 K under real conditions of connection of the bimetallic pipes in the cryomodule at a high RF of 3.9 GHz. JINR and the FERMILAB Accelerator Department worked out the strategy of the work and defined its fulfillment steps.

The high RF test is planned to be performed in the horizontal test system (HTS). For this purpose, the pipes were welded in pairs by their titanium ends in an argon atmosphere glovebox. Prior to that, all tubes were again subjected to thermal cycling and vacuum test at 77 K. The results of the Dubna and Pisa tests were confirmed. Preparation of cryogenic tests in the vertical test Dewar (VTD) began. The pipes were alternatively connected to the high-vacuum residual gas analyzer (RGA), and the leak rate measurement results were the same.

Tests of the pipes at the superfluid helium temperature (2 K) were conducted in the VTD and the HTS.

For VTD tests, the pair to be tested was connected to the RGA, evacuated, and immersed in superfluid helium.  In the HTS, the transition elements are tested in real conditions of superfluid helium with insulating vacuum around.

Pipe pair 5+6 was tested twice in the VTD for a short time of 0.3 h and 1.5 h and in the HTS for 30 h. Measurements demonstrated stable results: no leaks at the level of $<(2-3) \cdot 10^{-11}$ Pa·m$^3$/s. Pipe pair 9+11 from the last batch of pipes with changed sizes was tested in the VTD for 23 h, and pipe pair 5+6 was tested for 20 h. The result was quite impressive again: no leaks at the level of $Q<10^{-12}$ Pa·m$^3$/s [6].

## BIMETALLIC Nb+SS TRANSITION ELEMENT

Based on the experience gained with the Ti+SS samples, we got down to tackling the next problem within the ILC program, which was to investigate into a possibility of replacing the titanium shell of the helium cryostat with a stainless-steel shell, which would appreciably decrease the accelerator cost. This required development of a transition element from the stainless-steel shell to the niobium cavity, the main accelerating component of the accelerator [8].

In Sarov, the first four Nb+SS samples were made under two explosion welding schemes proposed by JINR, external explosion cladding (Fig. 5a) and internal explosion cladding (Fig. 5b).

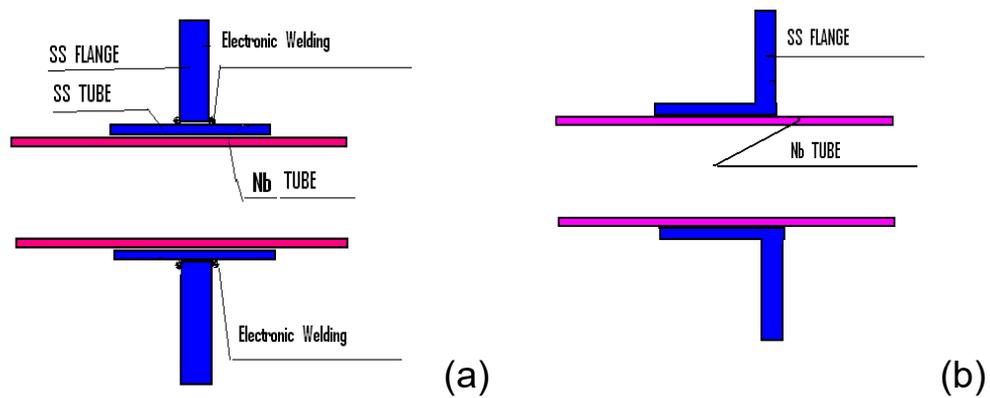

*Fig. 5. Schemes of welding together a Nb pipe and a stainless-steel flange.*

Apart from the initial task of making a Nb → SS transition element by explosion welding, we also faced an important problem of finding out whether explosion-welded Nb+SS transition elements could be used in ILC cryomodules of the fourth generation. The point was that in our sample the transition from the niobium cavity to the steel flange was through electron beam welding of the niobium pipe to the niobium cavity. The niobium melting temperature is 2460°C, and a question arises as to whether the Nb+SS joint is capable of withstanding this great thermal load. Preliminary tests of the niobium–stainless steel joint for tightness by putting one sample through thermal cycling in liquid nitrogen yielded quite satisfactory results: after six thermal cycles and ultrasound cleaning at the background leak detector reading of $2 \cdot 10^{-9}$ atm·cm$^3$/s no leak was found in the sample (He gas was blown in through the sample). The decisive test had to be a test of the welded joint for stability under high temperature, since the transition element operates when welded to the niobium cavity. Niobium rings with sizes corresponding to the size of the niobium pipe in the transition element were prepared for the test. They were welded to the pipe on both ends by electron beam welding at Sciaky Company (Chicago). The very first leak rate measurement at room temperature revealed large leak at three points on the Nb–stainless steel bond line (Fig. 6). This demonstrated the great effect produced by extreme temperatures on the tightness of the explosion-welded joint. The solution to the problem should be sought in structural changes of materials caused by explosion welding and extreme temperature variations [10–11].

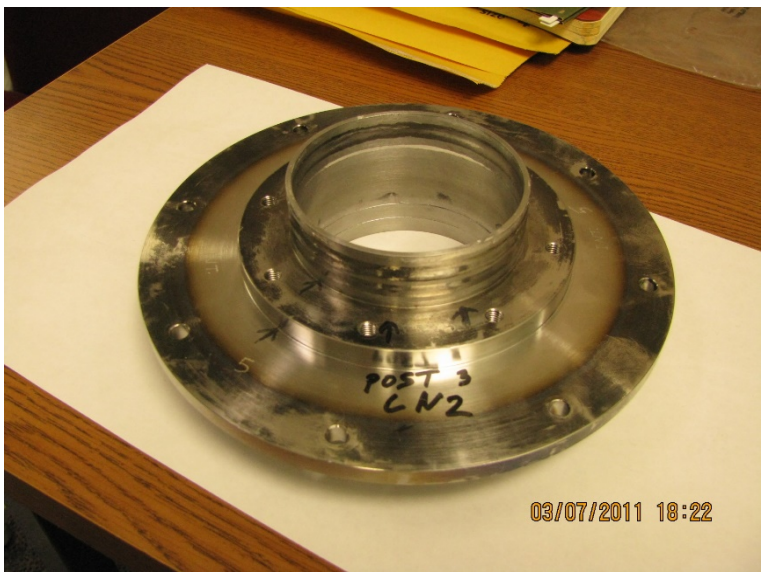

*Fig. 6. Nb+SS transition element with the found leak points.*

It is known that mechanical treatment (hammering, rolling, welding) of metal articles may result in residual stresses inside the material or the article caused by its plastic deformation. Breaking of mechanical components or structures is consequence of not only operation-caused stresses but also superposition of these stresses and residual stresses. We used the neutron diffraction method to measure residual stresses. Neutrons as a probe of the atomic structure have a substantial advantage over X rays and electrons because neutrons interact with atomic nuclei, which may result in their diffraction. We measured residual deformation in explosion-welded bimetallic Ti+SS pipes. They were chosen because the welding process is identical for the Ti+SS and Nb+SS joints, and thus the physics of diffusion in the materials during explosion and occurrence of residual stresses at their welded joints should also be identical. The measurements were carried out in the neutron beam from the ISIS reactor at the Paul Scherrer Institute (Switzerland) using the POLDY stress diffractometer [7]. The results of measuring residual stresses in the bimetallic Ti+SS pipe while scanning the titanium-to-stainless steel joint is shown in Fig. 7.

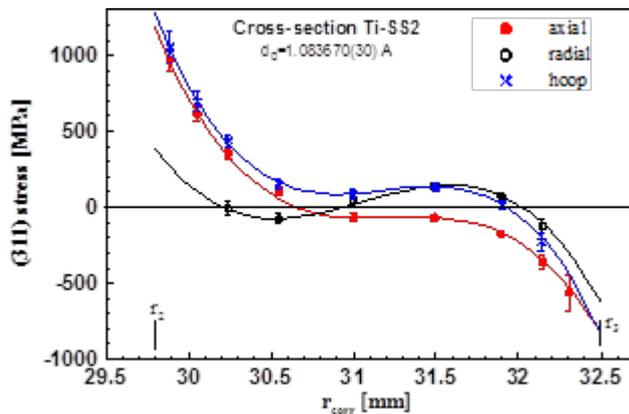

*Fig. 7. Measured (points) and fitted (curves) radial dependences of the stress tensor components for the peak (311) in the Ti-SS cross section.*

The residual stress reaches quite an appreciable value of ~1000 MPa. During electron beam welding or deep cooling in liquid helium additional thermal stresses arise, and their superposition may cause transition of titanium to the state of deep plastic region. This may give rise to local microcracks in the Ti+SS (or Nb+SS) welded joint. The Nb+SS samples were thermally tempered to relieve residual stresses both before and after electron beam welding. The thermal tempering procedure included heating in vacuum to the temperature of 750°C at a rate of 3°C/min, keeping at this temperature for 120 min, and natural cooling in vacuum for a night with the thermal oven switched off. After cooling, the samples underwent ultrasound cleaning in a 2% solution of the Micro90 cleaner in deionized water at room temperature for 30 min. The leak test was conducted by thermal cycling in liquid nitrogen; a Dupont (Ametek upgrade) leak detector with the sensitivity of $10^{-10}$ atm·cc/s was used, which was calibrated with an article having a standard leak rate of $5.3·10^{-8}$ atm·cc/s. A total of six thermal cycles were performed. The results were minimally different from cycle to cycle. The table lists the results for one of the cycles.

| Cycle 2 | Sample 1 | Sample 2 |
|---|---|---|
| Background leak at room temperature | $1.0·10^{-9}$ atm·cc/s | $5.3·10^{-9}$ atm.cc/s |
| Background leak at 77 K | No change | No change |
| Test article in polyethylene bag, He gas injected, 77 K | $1.2·10^{-9}$ atm·cc/s | $5.0·10^{-9}$ atm.cc/s |
| Test article heated to room temperature, He gas injected | No change | $4.9·10^{-9}$ atm.cc/s |

The next step was a test at the helium temperature of 2 K. The test sample was fixed in a special rig to be inserted to the VTD with liquid helium (Fig. 8). The cooling process was 15 to 10 h long, which included cooling with liquid nitrogen, injection of two-phase helium and cooling to 4 K, and evacuation and transition of helium to the superfluid state at 1.6–2 K. Mass distributions of the gases remained after the evacuation to $10^{-9}$ Torr were measured by the RGA for an hour at the temperature of 2 K, and no leak was observed at the background level of $4.6 \cdot 10^{-9}$ atm·cc/s.

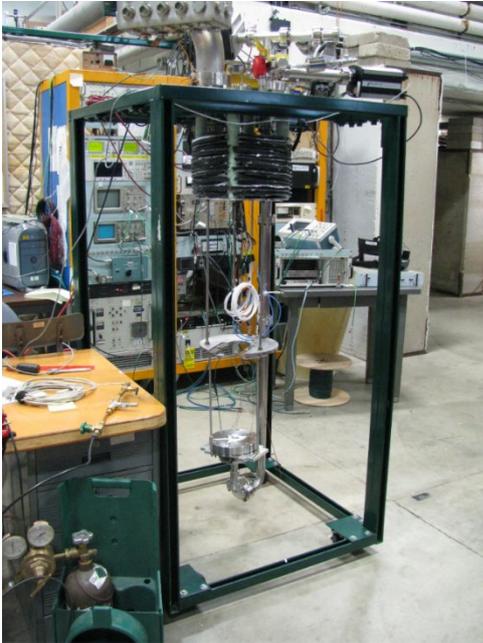

*Fig. 8. Assembled test sample ready to be inserted into the VTD.*

## MODIFICATION OF THE Nb+Ti+SS TRANSITION ELEMENT

To avoid the effect of residual stresses on the quality of the Nb+SS transition element, a way had to be found for joining dissimilar materials so that extreme thermal and mechanical loads did not affect the tightness and strength of the joint. It was also necessary to cope with the effect produced on the parameters of the transition elements by the difference in the linear expansion coefficients of their different component metals arising from huge temperature variations. The solution was found in discussions with a group of specialists from the world-known Paton Electric Welding Institute (Kiev, Ukraine).

It is known that the best-quality welded joints are obtained for similar materials, e.g., niobium and niobium or stainless steel and stainless steel. In our case, the transition element consists of at least two metals, niobium and stainless steel, and thus any fusion welding process, including electron beam welding, is unacceptable due to formation of intermetallides like $Nb_xFe_y$, which prevent obtaining the required tightness of the transition elements.

Earlier experiments showed that electron beam welding of niobium and titanium did not result in formation of intermetallides and proved the necessary helium and vacuum tightness of the joints. Thus, the following process was proposed for making a transition element.

A stainless-steel disc is clad with titanium on both sides by explosion welding. After this trimetallic piece is given a necessary shape (by leveling and turning to size), a hole for the niobium pipe is cut. The pipe is inserted in the hole and welded to titanium by electron beam welding (Fig. 9).

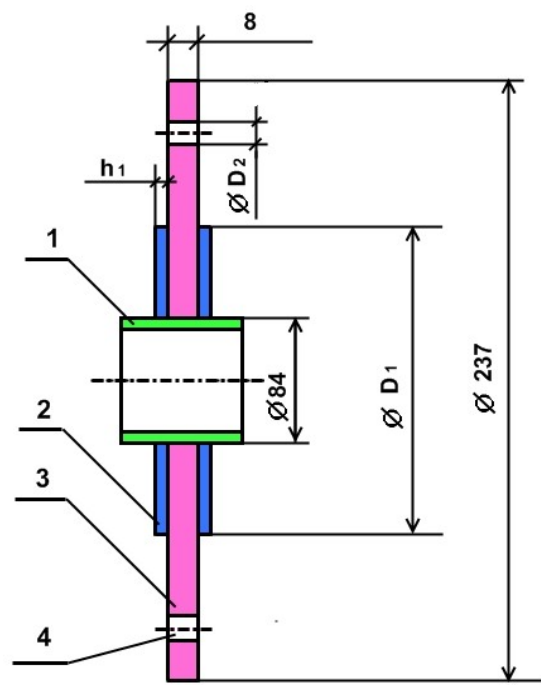

Fig. 9. Transition element design preventing formation of niobium intermetallides during welding: (1) niobium pipe, (2) titanium plates, (3) stainless-steel disc

Advantages of the proposed process are as follows:
- helium tightness is ensured by welding niobium and titanium that are easy to weld;
- the hole in the flange is made to the size of the niobium pipe, and it is possible to weld in the cavity pipe instead of the transition pipe;
- possible formation of intermetallides in the explosion-welded titanium–steel joint does not affect the operating capability of the transition element since helium cannot get into the niobium pipe through it;
- technologically, explosion welding of flat pieces is considerably simpler than welding of tubular pieces and allows producing joints with the maximum possible stability of quality, which decreases the defective work probability;
- in the case of defects after explosion welding, cheaper steel–titanium workpieces will be rejected;
- if necessary, the steel–titanium flange can be thermally treated in a normal (nonvacuum) oven to reduce residual stresses;
- expenditure of steel and niobium decreases.

Two trimetallic Nb-Ti-SS transition elements were produced by this scheme at the Paton Electric Welding Institute [12-13]: their flanges were made of 10X17H13M2E (316L) steel, and the pipes were of **Nb RRR 300** niobium (Fig. 10).

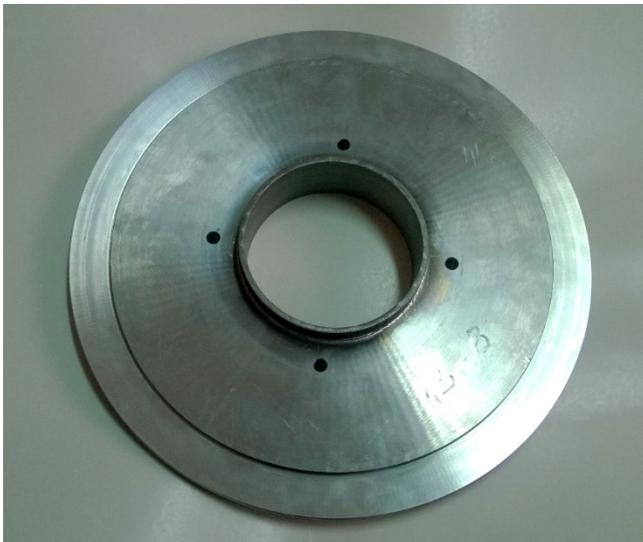

Fig. 10. Nb+Ti+SS transition element.

The quality of the explosion-welded titanium–steel joint was evaluated using the standard bend test, layer tear test, and layer shear test. Figure 11 shows the test sample after the bend test. Bent at the angle of 180°, the sample kept its integrity, and no layering occurred. It is a rather sever test, and if the welding is of poor quality, the bond line is broken.

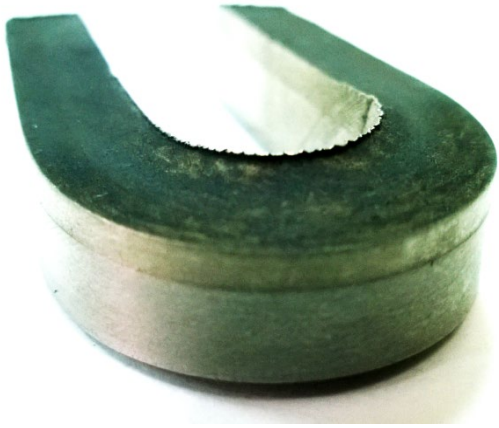

Fig. 11. Bimetallic steel–titanium sample after the bend test.

Figure 12 presents the scheme of the bimetal layer separation test.

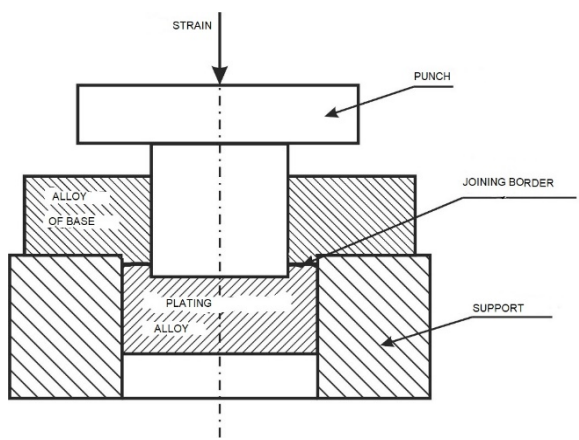

Fig. 12. Scheme of the layer separation test.

The samples were broken along the steel–titanium interface, which is typical of this pair of metals. The tensile strength was 375 MPa. The tensile test of the titanium sheet in the initial state showed that the yield point was 390 MPa, and the breaking point was 430 MPa.

The layer shear tests (Fig. 13) showed the strength to be at a level of 350 MPa. This high shear strength comparable with the tear strength is due to the wavy steel–titanium bond line.

Operation conditions of the transition element do not imply loads leading to layer shear or tear, and therefore the bond strength can be considered satisfactory.

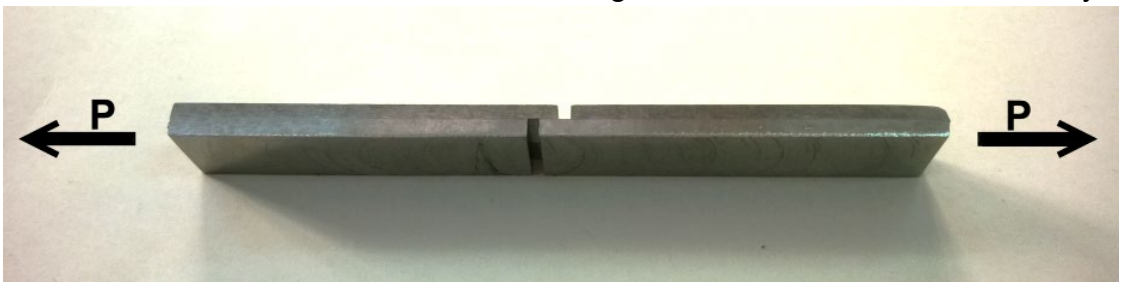

Fig. 13. Scheme of the layer shear test.

The welded bond strength was investigated using the Vickers hardness test. The results of hardness measurement under a load of 100 g are presented in Fig. 14.

It is seen in Fig. 14 that titanium and steel were considerably hardened by explosive impact. Titanium has its initial hardness as far as 300 μm away from the boundary, while steel is hardened deeper.

Thus, the investigations obviously show that the devised process for welding a trimetallic transition element is almost optimal. To gain larger statistics, the JINR-FNAL-INFN collaboration agreed to fabricate ten trimetallic Nb+Ti+SS samples at the Paton Electric Welding Institute and comprehensively test them in Kiev, Dubna, and Pisa/Genova.

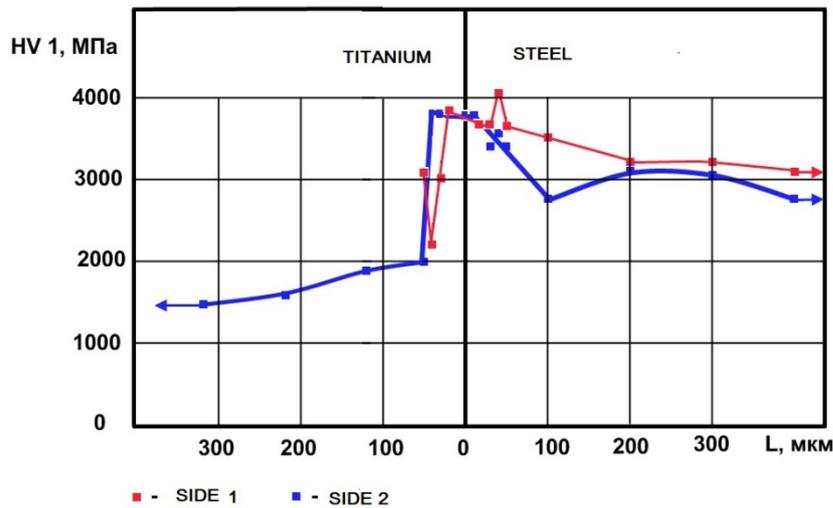

Fig. 14. Microhardness at the boundary of the explosion-welded steel–titanium bond.

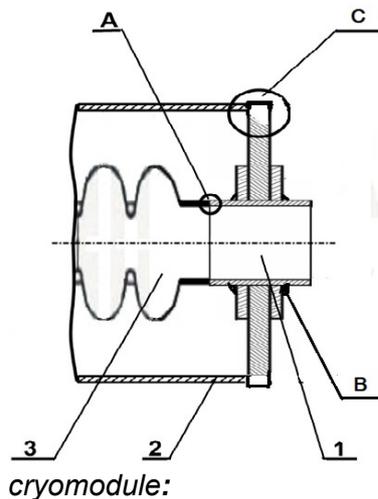

Figure 15 shows a fragment of the cryomodule to be connected to the trimetallic Nb+Ti+SS transition element. Positions **A** (Nb+Nb) and **B** (Ti+Nb) are welded by electron beam welding, and position **C** (SS+SS) is welded by ordinary electric arc welding.

Fig. 15. Joining of the Nb+Ti+SS transition element with the cryomodule:
1- Nb tube;
2- SS shell of cryomodule vessel
3- niobium cavity

To imitate position **A**, niobium rings of the corresponding diameter and 20 mm long were welded to the ends of the niobium pipe by electron beam welding. Tightness tests were performed at INFN (Pisa/Genova, Italy). To this end, a test device (Fig. 16) was built, which was reliably sealed with indium gaskets for tests at the liquid helium temperature of 4.2 K.

In Pisa, leak-tightness tests were carried out by thermal cycling at the liquid nitrogen temperature of 77 K using a VARIAN979 leak detector with the vacuum of $10^{-4}$ Torr and sensitivity of $0.1 \cdot 10^{-10}$ atm·cc/s.

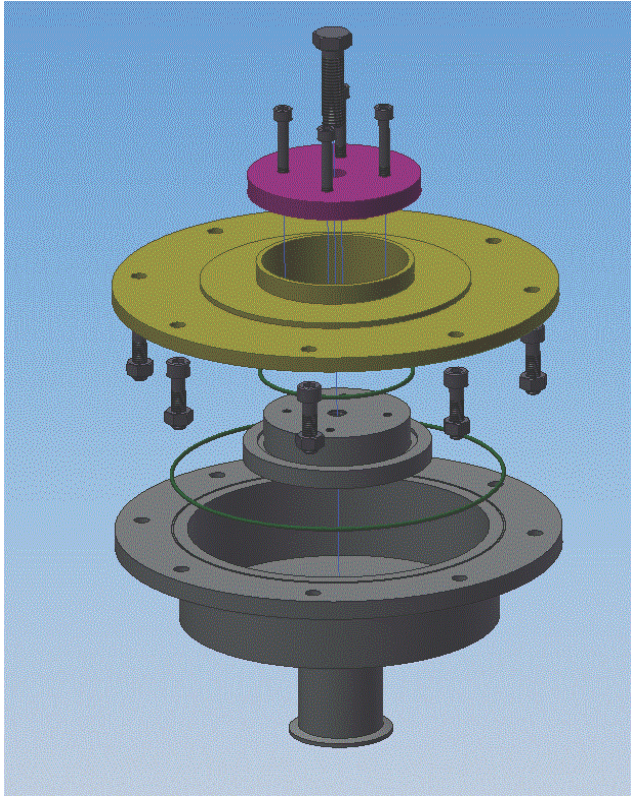

*Fig. 16. Test device.*

The tests were carried out using the proven scheme of thermal cycling in liquid nitrogen. The Nb+Ti+SS test sample was fixed in the test device, which was evacuated to $10^{-4}$ Torr and inserted in a Dewar flask with liquid nitrogen. Cooling of the test sample down to 77 K was indicated by cessation of nitrogen "boiling". Then the sample was taken out and heated to room temperature by heat guns. This cooling–heating cycle was repeated six to seven times. The background leak rate was measured at room temperature. For the leak test with the He gas, the test sample was placed in a plastic bag, the bag was tightly closed and punctured with a needle, and the He gas was blown in through the punctured hole. Tests with liquid helium were carried out at the liquid-helium setup of INFN (Genova). To save expensive helium and time for cooling the setup to the helium temperature, a device was made, in which five Nb+Ti+SS samples could be assembled at a time (Fig. 17) and then inserted in the Dewar flask and cooled simultaneously.

Achievement of the liquid helium temperature was controlled by temperature sensors attached to the upper and lower test samples. The inlet neck of the Dewar flask 80 cm in diameter was at the floor level, and the 4-m-high flask was dug into the floor and filled with liquid helium.

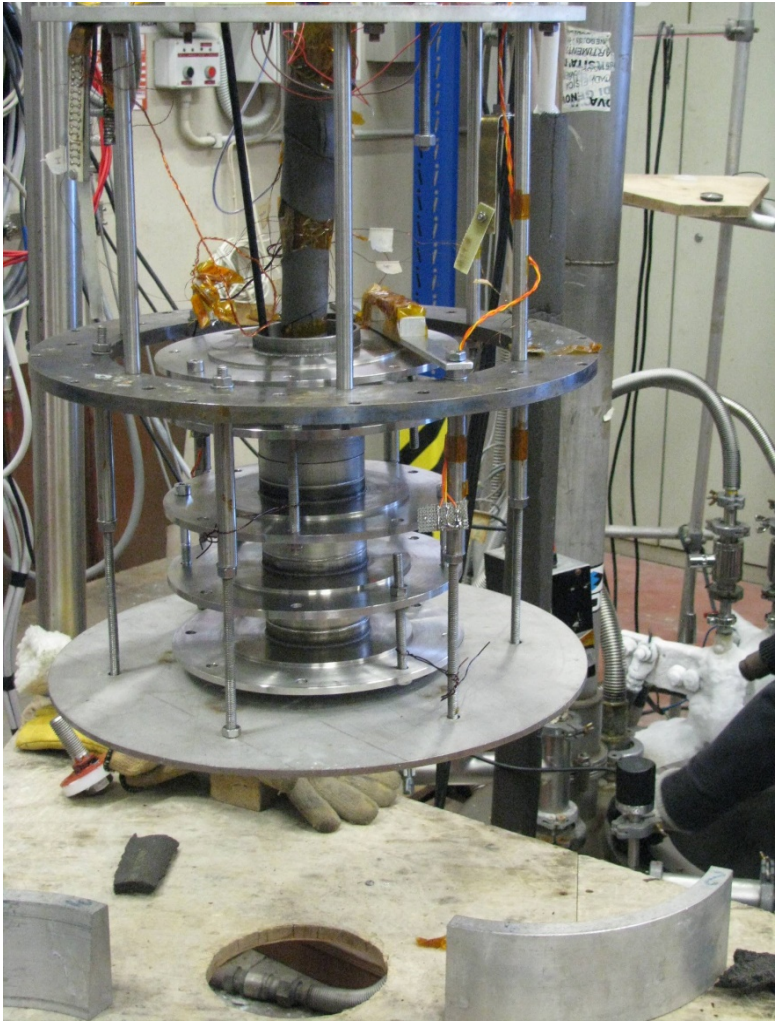

Fig. 17. Assembly of five samples before insertion in the Dewar flask with helium.

The filling process is about an hour long, and when 4.2 K is reached on both temperature sensors (Fig. 18), the cooling is stopped. Heating is approximately as long as cooling. The test samples are removed from the Dewar flask heated to room temperature, and cooled again. To save time and expensive gas, we performed only two thermal cycles with liquid helium.

Fig. 18. Readings of the upper (left) and lower (right) temperature sensors.

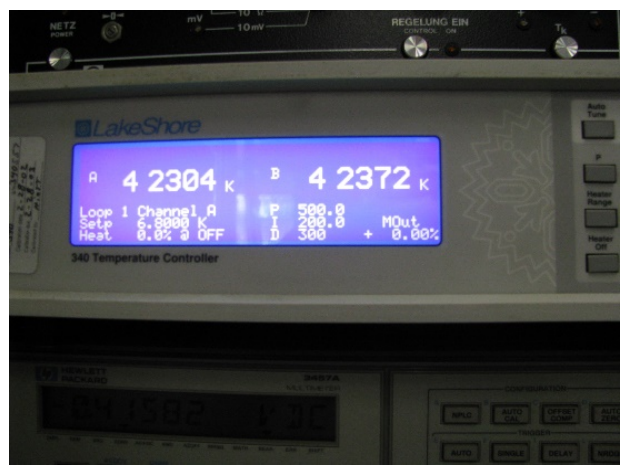

Table 1 presents test results for all 10 Nb+Ti+SS transition elements tested by thermal cycling with liquid helium [12-13].

# TABLE 1.

## LEAK TEST RESULTS AT LIQUID HELIUM THERMOCYCLES

| | LEAK RATE ($10^{-10}$ atm*cc/s), at 300K | | | |
| --- | --- | --- | --- | --- |
| | before LHe thermocycles | | after LHe thermocycles | |
| No. | background | He gas | background | He gas |
| 1 | 0.2 | 0.3 | 0.2 | No variation |
| 2 | 0.2 | No variation | 0.1 | No variation |
| 3 | 0.5 | 0.2 | 0.1 | No variation |
| 4 | 2.0 | 0.2 | 0.1 | No variation |
| 5 | 0.2 | 2.1 | 0.7 | 3.0 |
| 6 | 0.1 | 0.3 | 0.1 | 0.7 |
| 7 | 0.9 | 1.5 | 0.1 | 0.2 |
| 8 | 0.4 | 1.0 | 0.3 | 1.0 |
| 9 | 0.1 | 1.7 | 0.3 | 0.2 |
| 10 | 0.1 | No variation | 0.2 | 0.6 |

As is evident from the table, all transition elements meet the requirements on the stable operation of the cryomodule. The developed technology for explosion welding of dissimilar materials ensured reliably tight and strong connection of cryomodule components.

# CONCLUSIONS

The results reported in this work are final for the series of investigations into possible application of a unique explosion wielding technology for making elements of cryogenic systems operating under extreme conditions of ultralow temperatures. To replace titanium with stainless steel as a material for the shell of the cryostat for liquid helium (1.8 K), an entirely novel transition element has been designed for connecting the niobium cavity to the stainless-steel shell. Physicochemical properties of this unique module for the linear accelerator are extremely sensitive to tolerances on the design parameters. Therefore, high-quality technology development work was done to ensure both close adherence to the tolerances and stability of the product output characteristics against tolerance variations unavoidable in large-scale quantity production.
 Our results have shown that the explosion welding method allows manufacturing quite adequate bimetallic and trimetallic components of cryogenic parts for linear accelerators. It was also shown that residual stresses resulting from the explosion welding and their use at extremely low and high temperatures are quite large (≈1000 MPa), which may cause plastic deformation and break the bond. Explosive joining of niobium and stainless steel through intermediate bonding titanium led to a fundamentally important result: for the first time a trimetallic Nb+Ti+SS joint was produced for the ILC cryomodule. Development of the Nb+Ti+SS joining process greatly simplifies the design of the fourth-generation cryomodule: the liquid-helium lines and the cryostat (with the Nb cavity inside) are made of stainless steel instead of titanium, which noticeably reduces the cryomodule cost. Preliminary leak tightness tests of the Nb–stainless steel joint by thermal cycling in liquid nitrogen and liquid helium at the background rate of $0.2 \cdot 10^{-10}$ atm·cm$^3$/s on the helium leak detector did not reveal a leak (the test sample was blown through with the He gas), which indicates a high tightness and strength quality of the joint.

Our proposed design rules out influence of residual stresses on the joint quality, leakage of the Nb+Ti+SS joint, and effect of the difference in linear expansion coefficients of the assembly components. At present, explosion welding is the single option for making Nb+SS transition elements. This method is also be applicable to any other cryogenic systems.

## ACKNOWLEDGEMENTS

We would like to extend our great appreciation to our co-authors professor A. Balagurov from FLNP, JINR (Dubna); Dr. A. Venter from PSI, (Switzerland); I. Malkov, V. Perevozchikov from RFNC-VNIIEF (Sarov); N. Dhanaraj, M. Foley , R. Kephart, A. Klebaner, D. Mitchell, and W. Soyars from FNAL (Batavia, USA) for their invaluable contribution to the current publications in the course of the work.

We express our thanks to     K.A. Anderson, H. Carter, L. Cooley , R. Montier,  D. Plant, B. Smith, B. Tennis and V. Yarba    from Cryogenic Department of the Accelerator and Technical Divisions of FNAL for the organization and technical assistance in the preparation and conduct of the tests, as well as to the technical personnel of the INFN for construction of the testing facility and assistance in conducting the tests in Pisa.

We would like to express our special gratitude to the technical personnel of the RFNC-VNIIEF for their fruitful creative work on the development of the unique bimetallic pipe joints.

We are also deeply grateful to the former Director of the JINR Academician A.N. Sisakian, scientific supervisor from the RFNC-VNIIEF Academician R.I. Ilkaev, former Director of the FERMILAB Professor Pier Oddone, Director of the E. Paton Electrowelding Institute Academician B. Paton, and Directorate of the INFN Pisa for their far-sighted initiative, which has led to a quite promising approach of using the explosion welding method for modernization of the cryosystem of the International Linear Collider (ILC).  To achieve success in the development of the aforementioned approach was only possible under their constant support.

# ABSTRACT

Results of testing modified components for the cryomodule of the International Linear Collider (ILC) are summarized. To reduce the ILC project cost, it is proposed to replace titanium cryomodule components with stainless steel (SS) ones. New bimetallic transitions (Ti/SS, Nb/SS) have been produced by a unique method based on explosion welding. Successive upgrading of these components to the latest version of the Nb/Ti/SS transition element has led to improvement of the ILC cryomodule. This new component resolves problems of residual stress, and its specific design prevents the possibility of a shift due to the difference in the linear expansion coefficients of the constituent metals. Leak tests with the He gas revealed no leaks at the background rate of ≈$0.2 \cdot 10^{-10}$ atm·cc/s. The test results are very encouraging. The up-to-date design of trimetallic Nb/Ti/SS element promises technologically simpler and less expensive manufacture. Investigations have shown that explosion welding allows unique trimetallic components to be made not only for cryogenic units of accelerators but also for laboratory equipment and for general engineering applications.